# A PORTABLE POTENTIOMETRIC ELECTRONIC TONGUE LEVERAGING SMARTPHONE AND CLOUD PLATFORMS

*Patrick W. Ruch[a], Rui Hu[a], Luca Capua[a], Yuksel Temiz[a], Stephan Paredes[a], Antonio Lopez Marin[b], Jorge Barroso Carmona[b], Aaron Cox[b], Eiji Nakamura[c], Keiji Matsumoto[c]*

[a] IBM Research – Zurich, Säumerstrasse 4, 8803 Rüschlikon, Switzerland
[b] IBM T.J. Watson Research Center, Yorktown Heights, NY 10598, USA
[c] IBM Research – Tokyo, Kawasaki, 212-0032, Japan

## ABSTRACT

Electronic tongues based on potentiometry offer the prospect of rapid and continuous chemical fingerprinting for portable and remote systems. The present contribution presents a technology platform including a miniaturized electronic tongue based on electropolymerized ion-sensitive films, microcontroller-based data acquisition, a smartphone interface and cloud computing back-end for data storage and deployment of machine learning models. The sensor array records a series of differential voltages without use of a true reference electrode and the resulting time-series potentiometry data is used to train supervised machine learning algorithms. For trained systems, inferencing tasks such as the classification of liquids are realized within less than 1 minute including data acquisition at the edge and inference using the cloud-deployed machine learning model. Preliminary demonstration of the complete electronic tongue technology stack is reported for the classification of beverages and mineral water.

*Index Terms*— Electronic tongue, polymer, electrodeposition, smartphone, cloud computing

## 1. INTRODUCTION

Electronic tongues can distinguish complex liquids by combining cross-sensitive sensor arrays with machine learning. In contrast to conventional analytical techniques for chemical fingerprinting, such as mass spectrometry for food authentication [1], electronic tongues based on electrochemical sensing principles may be miniaturized for use in portable or remote sensing [2]. Such compact analytical devices enable chemical fingerprinting at the point of use, thereby reducing time and effort related to sample handling and transport. Yet, there are at present no commercial electronic tongues available in a form factor suitable for portable and remote chemical fingerprinting [3]. In the present contribution, a technology framework is presented for the development and deployment of miniaturized electronic tongues based on potentiometry. The hardware and software components of this framework are presented along with preliminary results related to classification of beverages and mineral water using a handheld device, smartphone and cloud computing services. The technology stack is expected to support the development and deployment of electronic tongues for chemical fingerprinting in portable or remote applications.

## 2. METHODOLOGY

### 2.1. Potentiometric sensor arrays

Substrates for the sensor arrays were either oxidized silicon Si (0.5 mm) / $SiO_2$ (500 nm) or conventional printed circuit boards (PCBs) of thickness 1.6 mm. Silicon substrates were metallized with sputtered Ti (10 nm)/Pt (100 nm) electrodes patterned by optical lithography and selectively masked with photopatterned SU-8 polymer to expose four circular pads to sample liquid. PCBs were ordered with standard electroless nickel immersion gold (ENIG) metallization defining four circular pads for contact with sample liquid. The electrode pads with Ø2 mm were coated by polypyrrole (PPy) via electropolymerization of pyrrole (Py), whereby different deposition conditions and precursor solutions were chosen to yield different ion sensitivities. As-deposited PPy from chloride-containing electrolytes yields films with anionic sensitivity, e.g. toward $Cl^-$ (Figure 1a). As found previously by Michalska *et al.* [4], incorporation of hexacyanoferrate(II) (FOCN) anions in PPy with appropriate concentrations of FOCN and Py in the precursor solutions and suitable electrodeposition conditions leads to cationic sensitivity of the resulting films (Figure 1b). Preliminary classification experiments were performed by functionalizing each electrode in one sensor array by PPy films exhibiting different ion selectivity and sensitivity, mainly toward $K^+$, $Na^+$ and $Cl^-$, according to Table 1. Sensors 3 and 4 were nominally identical for testing purposes.

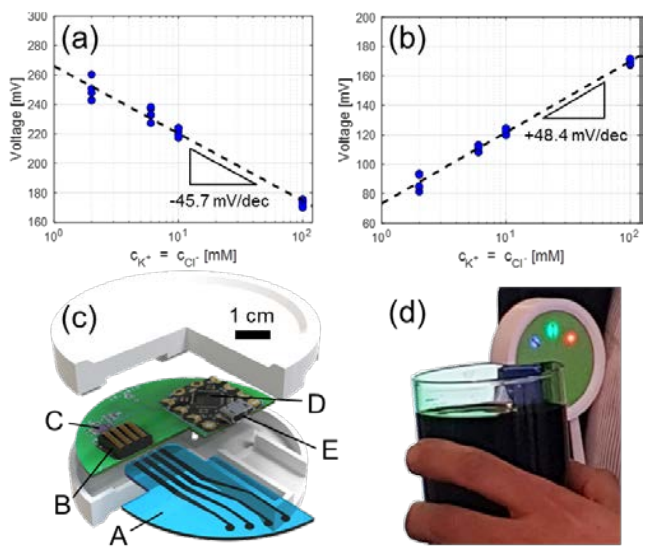

**Figure 1** (a) Anionic sensitivity of electrodeposited PPy films with change in KCl concentration, (b) cationic sensitivity of electrodeposited PPy/FOCN films with change in KCl concentration, (c) assembly of sensor array on PCB [A] with spring-loaded connector [B] and read-out electronics comprising an ADC [C] and microcontroller [D] with micro-USB port [E] for connection to smartphone, and (d) photograph of packaged electronic tongue device immersed in liquid sample.

**Table 1** Potentiometric sensors deposited on electrode array.

| ID | Precursor solution | Electrodeposition conditions |
|---|---|---|
| 1 | 0.1 M Py<br>0.1 M $CaCl_2$ | 0.8 V vs. SCE<br>60 s |
| 2 | 0.2 M Py<br>0.05 M $K_4Fe(CN)_6$ | -0.5 V...0.8 V vs. SCE @ 20 mV/s<br>3 cycles |
| 3, 4 | 0.1 M Py<br>0.1 M $K_4Fe(CN)_6$ | 0.8 V vs. SCE<br>60 s |

### 2.2. Data acquisition board

In lieu of an external reference electrode, the measurement of differential voltages across the four terminals of the sensor chip (A in Figure 1c) as a function of time was chosen as sensor signal for discriminating liquids. This allowed a single spring-loaded connector (contact pitch 2.54 mm, B in Figure 1c) to be used for interfacing of the sensor chip with the data acquisition (DAQ) board. The DAQ comprised a 16-bit 4-channel analog-to-digital converter (ADC, TI ADS1115, C in Figure 1c) with programmable range and resolution set to ±2 V and 0.0625 mV, respectively. The potentiometric data was transmitted from the ADC to an Arduino microcontroller (DFrobot Beetle, D in Figure 1c) using the $I^2C$ protocol. Finally, a micro-USB connector (E in Figure 1c) was used to transfer data to a smartphone via USB On-The-Go (OTG).

### 2.3. Portable sensor package

Housings for the DAQ board were 3D-printed with windows for LED status lights and slots for connecting the sensor chip and USB-OTG cable. The circular design of the assembled device allows self-supported clamping onto a glass containing the sample liquid (Figure 1c, d).

### 2.4. Machine learning model deployment

An overview of the present framework for calibration data generation and model training ("Training") as well as blind liquid sample testing and classification ("Inference") is given in Figure 2. Transient potentiometric sensor data was recorded for three beverages A-C (6-8 measurements per sample) and four still mineral waters I-IV (24 measurements per sample). Prior to insertion of the sensor array in the sample liquid, the sensor array was stored in 0.1 M KCl which provided stable starting potentials (Figure 3a, first 20 seconds).

Data processing involved subtracting the measured voltage differentials by their respective average values in the KCl solution, eliminating the data obtained while the sensor was in the KCl solution and finally concatenating the three differential voltages (Figure 3b). A supervised random forest classification algorithm was trained with the processed data using the scikit-learn library for Python. The training data was stored on a cloud server and the trained model was also deployed in the cloud using the IBM Watson Machine Learning service. Further, a Python server was set up to carry out automated data processing and to invoke the deployed model during sample testing.

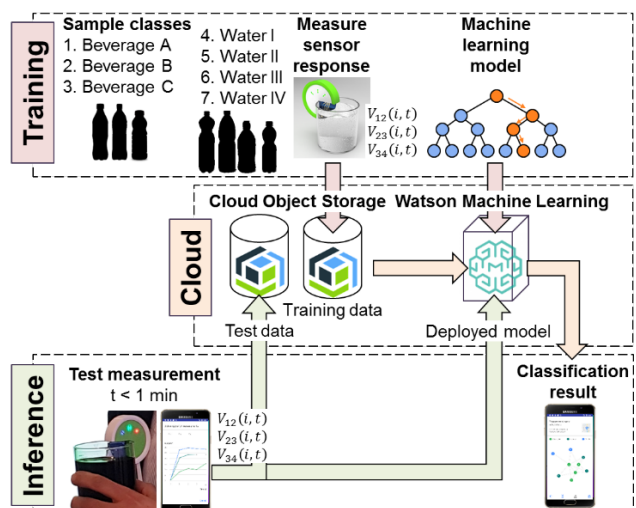

**Figure 2** Workflow for the portable electronic tongue device depicting stages for training and inference, along with cloud computing services for storing training and test data as well as execution of trained machine learning models during inference.

**2.5. Mobile app**

An Android app was developed to interface a smartphone (Samsung Galaxy A3, 2017) with the DAQ board described in section 2.2. The app displays the potentiometric data in real time and adds a date- and timestamp based on the smartphone's built-in time and location services. The raw test data is then transferred to a cloud server using Wi-Fi or cellular connectivity and stored in an Elasticsearch database. The Python server (section 2.4) processes the test data and sends the pre-processed data to the deployed classification model. Finally, the most likely classification result together with the corresponding confidence level and the pair-wise similarity score between the test and training data are sent back to the smartphone. In the mobile app, the classification results are visualized by means of a column graph showing the likelihood score by class, and the training and test data are shown in a force directed graph based on their pairwise similarity scores

## 3. RESULTS AND DISCUSSION

**3.1. Beverages**

Despite the small number of elements in the sensor array and predominant sensitivity toward small, monovalent ions, the classification of three soft drinks was achieved with an accuracy of 95.3% obtained by leave-one-out cross-validation on the training data set. Transient potentiometric features at short timescales (Figure 3b, index < 20) were found to be more important for successful classification than the final equilibrium potentials. This observation suggests that non-equilibrium potentiometry is a promising approach for rapid classification of liquids using portable or remote sensors.

**3.2. Mineral water**

Four brands of mineral water were used for training and testing with the same hardware as described above. Classification of these samples was achieved at an accuracy of 61.7%, which is significantly lower than for the beverages.

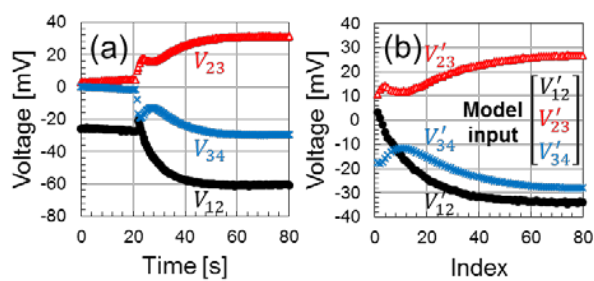

**Figure 3** (a) Raw differential voltage data and (b) processed data for training of machine learning model. The multi-array sensor was displaced from a 0.1 M KCl reference solution to the sample solution at 20 s, in this example a soft drink "beverage B".

**Table 2** Nominal concentrations of selected ions for the four studied brands of mineral water.

| Sample | Concentration [ppm] | | |
|---|---|---|---|
| | $Na^+$ | $K^+$ | $Cl^-$ |
| I | 4 | 2.5 | 16 |
| II | 7.3 | 4.9 | 3.7 |
| III | 6.5 | 1 | 6.8 |
| IV | 6 | 2.5 | 20 |

The lower accuracy is expected due the limited range of selectivity of the simple four-electrode array and the limited variation in concentration of dissolved monovalent ions at the ppm level (Table 2). However, the portable solution still performed substantially better compared to human testers, who were able to perform correct classification of the four mineral water brands with an accuracy of 30.6%. The confusion matrices of the two test data sets are shown in Table 3.

**Table 3** Confusion matrices for (a) beverages and (b) mineral water. Values represent number of observations.

(a) Beverages

| Predicted \ True | A | B | C |
|---|---|---|---|
| A | 7 | 1 | 0 |
| B | 0 | 7 | 0 |
| C | 0 | 0 | 6 |

(b) Mineral water

| Predicted \ True | I | II | III | IV |
|---|---|---|---|---|
| I | 16 | 2 | 3 | 1 |
| II | 7 | 7 | 4 | 6 |
| III | 1 | 6 | 15 | 2 |
| IV | 0 | 3 | 1 | 20 |

## 4. CONCLUSION

A platform has been developed for portable and remote applications of electronic tongues based on on-chip sensor arrays, a smartphone interface and cloud computing. The use of different or a greater number of sensing elements can be easily accommodated. Thus, an end-to-end solution for the deployment and testing of portable and remote electronic tongues is proposed. The incorporation of cloud computing services allows combining measurement data from multiple remote devices and the deployment of new machine learning models without modifying the device or software at the edge.